# Effects of photon escape on diagnostic diagrams for H II regions

C. Giammanco[1], J.E. Beckman[1,2], and B. Cedrés[1]

[1] Instituto de Astrofísica de Canarias, C. Vía Láctea s/n, 38200, La Laguna, Tenerife, Spain
e-mail: `corrado@ll.iac.es`
[2] Consejo Superior de Investigaciones Científicas, Spain

**Abstract.** In this article we first outline the mounting evidence that a significant fraction of the ionizing photons emitted by OB stars within H II regions escape from their immediate surroundings, i.e from what is normally defined as the H II region, and explain how an H II region structure containing high density contrast inhomogeneities facilitates this escape. Next we describe sets of models containing inhomogeneities which are used to predict tracks in the commonly used diagnostic diagrams (based on ratios of emission lines) whose only independent variable is the photon escape fraction, $\xi$. We show that the tracks produced by the models in two of the most cited of these diagrams conform well to the distribution of observed data points, with the models containing optically thick inhomogeneities ("CLUMPY" models) yielding somewhat better agreement than those with optically thin inhomogeneities ("FF" models). We show how variations in the ionization parameter $U$, derived from emission line ratios, could be due to photon escape, such that for a given region from which 50% of its ionizing photons leak out we would derive the same value of $U$ as for a region with no photon escape but with an input ionizing flux almost an order of magnitude higher. This effect will occur whether the individual inhomogeneities are optically thick or thin. Photon escape will also lead to a change in the derived value of the radiation hardness parameter, and this change differs significantly betwen models with optically thin and optically thick clumps. Using a rather wide range of assumptions about the filling factor of dense clumps we find, for a selected set of regions observed in M 51 by Díaz et al. (1991) an extreme limiting range of computed photon escape fractions between near zero and 90%, but with the most plausible values ranging between 30% and 50%. We show, using oxygen as the test element, that models with different assumptions about the gas inhomogeneity will tend to give variations in the abundance values derived from diagnostic diagrams, but do not claim here to have a fully developed set of diagnostic tools to improve abundance determinations made in this way. We do present an important step towards an eventual improvement in abundance determinations: the combination of line ratios with the absolute H$\alpha$ luminosity of a given H II region, which allows us to determine the photon escape fraction, and hence resolve the degeneracy between $U$ and $\xi$. We use observational data of this type show that a large set of H II regions in M 101 observed by Cedrés and Cepa (2002) all show significant photon escape with values of $\xi$ ranging up to 60% in the "leakiest" cases.

## 1. Introduction

Diagnostic diagrams for H II regions use emission line intensities and ratios in order to measure their basic physical parameters, motivated principally by the desire to use emission lines to derive reliable values of element abundances. In a recent article (Giammanco et al., 2004) (paper I.) we discussed the propagation of ionizing radiation through H II regions, with emphasis on the effects of optically thick density inhomogeneities on this propagation, and bringing out some of the effects on the interpretation of selected line ratios. In the present article we will concentrate on the specific need to take into account the escape of ionizing photons from H II regions when interpreting observations of line strengths and ratios. In the models of H II regions discussed in paper I we remarked that inhomogeneous H II regions, whether the clumps are optically thick or optically thin, present lower capture cross-section (ie. higher escape probability) for ionizing photons than homogeneous regions with the same gas mass. Here we will quantify this and derive some of the effects on specific diagnostic diagrams which have been widely used in the literature.

The importance of the escape of ionizing photons from H II regions has been increasingly understood in recent years. Since the importance of the diffuse ionized component of the interstellar medium (the 'DIG') was realized, (Reynolds et al., 1974) its possible origin in photons originating in OB stars was postulated, with the consequent implication that these photons must be able to escape from their local surroundings, the H II regions. Dove et al. (1994) presented a quantitative model in which the porosity of the regions and of the DIG itself permitted the Lyman continuum photons to be transmitted from their points of origin to points of ionization over 1 kpc away in spite of the high cross-section of atomic hydrogen for ionization, and this offered at least a partial response to those who had suggested that the presence of significant quantities of ionized

*Send offprint requests to*: C. Giammanco

gas at these distances from the nearest OB association required an alternative ionization mechanism (e.g. Sciama, 1995).

Although there is evidence, comparing emission line ratios in the DIG with their counterparts in H II regions, that a fraction of the ionization of the DIG may well be caused by shocks or turbulence (Collins & Rand, 2001; Hoopes & Walterbos, 2003), it is also clear from these studies that the major fraction of this ionization most probably comes from the OB stars in H II regions. A similar conclusion was drawn by Zurita et al. (2002), from a purely morphological study of face on galaxies, in which the pattern of surface brightness in H$\alpha$ was well reproduced using models in which defined fractions of the ionizing photons are escaping from the observed H II regions. The model, which used each measured H II region as a source of Lyman continuum, and took into account the observed H I surface density distribution of the galaxy under study (NGC 157) Ryder et al. (1998) gave a very good account of the H$\alpha$ surface brightness distribution of the DIG without recourse to any additional ionization sources, although the intrinsic uncertainties in the combination of observations and models do not exclude the presence of mechanical causes for a limited fraction of the observed DIG emission.

However for the purposes of the present article, all we wish to note is that the evidence that ionizing photons do escape from H II regions is clearly present in the articles cited above. In addition there is more direct evidence from individual H II regions that photons are leaking from H II regions. Oey and Kennicutt (1997) compared the Lyman continuum luminosities predicted from the OB stars within given associations in the LMC to that inferred from the observed H$\alpha$ luminosities of their surrounding H II regions and inferred that a fraction of the ionizing flux of up to 50% per region was escaping. In a detaileed study using photoionization modelling of the H II region NGC 346 in the SMC, Relaño et al. (2002) showed that around half the ionizing flux produced by the OB stars in this region must be escaping. For NGC 628, NGC 1232 and NGC 4258 Castellanos et al. (2002) found an escape fraction between 10%–73%. These studies also receive support, though less direct, from Zurita et al. (2000) who measured the total H$\alpha$ luminosities in the DIG of 6 external galaxies, and inferred that if the ionization of the DIG is indeed caused principally by Lyc photons escaping from H II regions, then close to 50% of these photons produced in the regions must be escaping. In addition Iglesias–Páramo & Muñoz–Tuñón (2002) show that the escape of photons is compatible with specific diagnostic diagrams, and Stasińska & Izotov (2003) compare models with diagnostic diagrams and suggest that the escape of ionizing photons may be increasing with the age of the H II regions.

The need to incorporate inhomogeneities in photoionization models of H II regions was noted even by Strömgren (1948) and models incorporating inhomogeneities were introduced by Osterbrock and Flather (1959). It is easy to see why a strongly inhomogeneous structure is implied by observations. Estimates of the mean electron densities in large H II regions, inferred from estimates of their emission measures (using e.g. H$\alpha$ surface brightness) and radii are of order $\gtrsim 1$ cm$^{-3}$ (Rozas et al. 1996) while in situ measurements of the electron density from emission line ratios give values of $\gtrsim 100$ (Zaritsky et al. 1994).

The most natural interpretation of this discrepancy is that the production of emission lines is strongly weighted towards the densest clumps of the ISM in an H II region, and that these are typically some two orders of magnitude denser than their surroundings. This conclusion rests on the physical basis that the strength of a recombination line is proportional to the square of the local electron density, so that the more tenuous parts of an H II region contribute very little to its net emission in these lines.

Since the early study of Osterbrock and Flather (1959) virtually all published research using H II region emission lines has used a "filling factor" approach to their interpretation. The filling factor $\phi$ is just the volume ratio of dense to tenuous gas, and to a first order approximation the contribution from the latter is set to zero. An implicit assumption in all of the models is that the clumps, though of relatively higher density, are optically thin, i.e. small enough to be fully ionized. However even a fairly small clump within a large H II region will not be optically thin. In Paper I (Fig. 1) we calculated the fraction of the volume of a clump of modelled radius 1 pc and density 100 atoms cm$^{-3}$ which would be ionized at varying distances from a central OB star cluster. We found that even in luminous H II regions, (with luminosities in H$\alpha$ of, say, more than $10^{38.5}$ erg s$^{-1}$) a clump with these properties would be optically thick at all radial distances greater than $\sim 0.1$ of the radial distance from the central cluster. Under these circumstances it makes sense to consider the effects of optically thick clumps on the radiative properties of the H II region. We compared a set of classical "filling factor" models with models containing opticlly thick clumps ("CLUMPY" models), in paper I, and one of the properties of the clumpy models which was clearly shown is that they entail photon escape from the region as a whole, while preserving apparent spectral signatures of ionization bounding.

In the present article we develop an approach to verification of hypothesis of leaky H II regions, by testing for compatibility with literature data on line emission. We confirm that photon escape is probably a property of most if not all regions, and go on suggest a method for estimating the photon escape fraction by measuring line ratios, and the absolute luminosity in H$\alpha$ of the region concerned. The first step is to define tracks in parameter space which mark the ensemble of those models for which only a single parameter is being varied. Then we can infer the key parameters which determine the properties of a given set of tracks, and select a range of values which is physically plausible for each of these parameters. The next step is to plot theoretically the evolution of the ionization parameter (a practical parameter which has been widely used in diagnostic diagrams, as explained in section 3) as a function of increasing photon escape fraction, from which we derived a semi-empirical method for estimating the value of the escape fraction from a given H II region, using measured emission line strengths and ratios. Comparing predicted tracks with observations on a set of diagnostic diagrams whose use in the literature is well established we show that the observational data are consistent with significant fractions of ionizing photon escape, and finally we consider how to modify the use of diagnostic diagrams to take this escape into good account.

## 2. The models and their parameters.

### 2.1. Definitions of the basic parameter used.

In our article (Paper I) we described a set of models dealing with the transfer of ionizing photons in CLUMPY media: 'CLUMPY" and 'Filling Factor" (FF) models, with optically thick and optically thin inhomogeneities respectively. Here we give a very brief summary of some of their more important parameters and assumptions. As input to both types of models we need to specify the luminosity and the spectral energy distribution of the ionizing sources, as well as the chemical composition of the H II region under test; the programs also require us to specify an inner and an outer radius for the region. Variation of the outer radius will effectively modify the escape fraction of the ionizing photons. In addition for the FF models, which contain optically thin clumps, we need to define a filling factor ($\phi$) while for the CLUMPY models, which contain optically thick clumps, we need to specify the absorption fraction per unit spherical shell layer within the H II region, the density and the sizes of the clumps which also determine the thickness of the concentric spherical shells into which the region is divided for computational purposes. For the geometrical deteils of the CLUMPY model we refer to section 2 of paper I. Here we summarize schematically the paremeters that we are to specifie for the costruction of a model:

- The Spectrum and luminosity of the ionizing source
- The internal and external radii of the H II region
- The element composition of the H II region
- The filling factor $\phi$ (for the FF models)
- The clump size and density (for the CLUMPY models)
- The absorption fraction f for shell (for the CLUMPY models)

### 2.2. Tracks and distances between them

In the models considered in the present article, the escape fraction, $\xi$, of the ionizing photons, is used as a free parameter, and it is convenient to define a *"track"* in parameter space by the set of models which differ only in the value of this escape fraction. Using these tracks will allow us to make comparisons between the different models and to test them against observations. To carry out these tests we will use a set of diagnostic diagrams whose details we will discuss below.
The method we will use to help us select the best models in situations where the observations are multivariate will be based on examining the distances between representative points and tracks in the diagnostic diagrams. These distances are defined using the classical definition of the distance between two points in a plane. If "x" and "y" are representative coordinates, we define the distance "d" between two points separated in x and y by $\Delta x$ and $\Delta y$ as $d \equiv \sqrt{\Delta x^2 + \Delta y^2}$. We can generalize this to define the distance D between two tracks in a diagnostic diagram where $x(\xi)$, $x_1(\xi)$ e $y(\xi)$, $y_1(\xi)$ are the coordinates of two models which belong to two different tracks but which have a common value of the ionizing photon escape fraction $\xi$. This is given by:

$$D^2 \equiv \int_\Xi \left( (x(\xi) - x_1(\xi))^2 + (y(\xi) - y_1(\xi))^2 \right) d\xi, \qquad (1)$$

where the range of integration $\Xi$ is the range of the escape fraction $\xi$ which we wish to sample. In our case for FF models $\xi$ varies between 0 and 90%, while for CLUMPY the range is from 30% to 50%. The lower limit for the CLUMPY models is not zero because in this kind of models the escape fraction is never zero. The definitions which we have presented are valid for a given diagnostic plane. However a model gives a good representation of the physical conditions in a volume of space (typically an H II region) if the predicted values lie close to the observed values in all the relevant diagnostic diagrams. It is not difficult to extend our definitions to cover a multidimensional space whose axes are mutually independent and which represents all the two-dimensional diagnostic diagrams which we want to use. We will therefore specify the space in which we are working whenever we calculate the distances between model track predictions or between model predictions and observations.

### 2.3. The parameters of greatest importance for our models.

In 2.1 we defined the most important parameters required to construct a model, and in 2.2 we defined a track and the distance between two tracks. On the basis of these choices and definitions we can quantify the relative importance of the parameters which are used in a given model. The idea is quite basic: the greater the separation between tracks generated by varying one of our parameters, the more important it is to use realistic values for that parameter. On the contrary, if by varying a given parameter we generate only a small separation between tracks, we can use a fixed value for that parameter without seriously affecting the comparisons between observations and models.

For both FF models and CLUMPY models we have made computations to examine whether major distances between tracks are produced when we vary the following parameters: the internal radius of an H II region, the input spectrum of ionizing radiation, and the element ('chemical') composition of the gas. We used as our reference plane the diagnostic diagram Log([O II]/[O II]) v. $r_{2,3}$ (for the definitions see Table 3).

For the FF models we did not, in practice, vary the filling factor, $\phi$. We fixed it at a value of $10^{-3}$, and used a local H I density "$n$" within the dense but optically thin clumps of 100 cm$^{-3}$, so that the mean density within the model as a whole takes a value of 3 cm$^{-3}$ (a canonical value taken from Rozas et al. (1996) who made observations of populations of H II regions in a number of galaxies). Varying $\phi$ or $n$ would change the ionizing parameters which govern the emission line strengths and ratios within a region, but the same effect is produced by varying the number of ionizing photons Q per unit of time.
Also in order to confine our parameter space to a manageable number of dimensions we have kept our clump size and the

| Change in value of inner radial limit from (0.06-6 pc) | D |
|---|---|
| black body 51230 K Orion abundances | 6.5 10$^{-4}$ |
| black body 51230 Solar abundances K | 3.6 10$^{-3}$ |
| WM-Basic | 2.3 10$^{-4}$ |
| **Different ISM composition (Solar or Orion)** | ... |
| Black body 51230 inner radius 0.06 pc | 3.9 10$^{-1}$ |
| Black body 41230 inner radius 6 pc | 4.6 10$^{-1}$ |
| **Different spectral energy distribution** | ... |
| Black body 51230-Black body 41230 | 2.5 10$^{-1}$ |
| Black body 51230-WM-Basic | 1.9 10$^{-1}$ |
| **Different stellar metallicities** | ... |
| WM-Basic 0.9-0.1 | 7.9 10$^{-3}$ |

**Table 1.** Parameter sensivity tests for FF models. We have compared pairs of models, measuring the distances between their tracks in the relevant parameter spaces, varying one parameter at a time while maintaining the others constant as shown in the table. In our treatment of these models in the text, we specify the value of a given parameter, and those parameters not so specified take their default values. For the FF models these are given by our reference model: internal radius 0.6 pc.; ISM abundances as in the Orion nebula; (using values given in HAZY, Ferland 1996, the manual for the use of his CLOUDY program suite), and filling factor 10$^{-3}$. In this reference model we took an ionizing spectrum with a pure black body form, at a temperature of 51230K, that of an O3 star, and used a total ionizing luminosity of 10$^{40.31}$ erg s$^{-1}$ which is the equivalent to that emitted by 10 O3. stars (using the results of Vacca et al. 1996). Note: although the units of the track separations are not specified here, this table and Table 2 are used to distinguish sensitive parameters, those with values >10$^{-1}$ from insensitive parameters, those with values <10$^{-2}$, so absolute values and units are not required. In the table we refere also to the WM-Basic stellar atmosphere (Pauldrach et al., 2001) with which we modelize 10 O3 stars, for more details see the text.

| Change in value of inner radius (4-6 pc) | D |
|---|---|
| Black body 51230 K Orion abundances | 5.1 10$^{-3}$ |
| Black body 51230 K Solar abundances | 4.3 10$^{-3}$ |
| WM-Basic | 5.6 10$^{-3}$ |
| **... (2-4 pc)** | ... |
| WM-Basic | 6.7 10$^{-3}$ |
| **Different ISM composition (Orion or Solar)** | ... |
| Black Body 51230 inner radius 4 pc | 3.6 10$^{-1}$ |
| **Different spectral Energy distribution** | ... |
| Black body 51230-Black body 41230 | 2.2 10$^{-1}$ |
| Black body 51230-WM-Basic | 9.0 10$^{-1}$ |
| **Different stellar metallicities** | ... |
| WM-Basic 0.9-0.1 (radios 4) | 7.9 10$^{-3}$ |

**Table 2.** Parameter sensitivity tests for CLUMPY models. The procedure for computing the track separations is the same as that used for Table 1, except for the default value for the inner radius which we have taken as 6 pc for reasons explained in the text.

| Symbol | Specific line or doublet |
|---|---|
| [O II] | [O II] $\lambda$ 3727 Å |
| [O III] | [O III] $\lambda\lambda$ 4959, 5007 Å |
| [S II] | [S II] $\lambda\lambda$ 6717, 6731 Å |
| [S III] | [S III] $\lambda\lambda$ 9069, 95032 Å |
| [N II] | [N II] $\lambda\lambda$ 6548, 6583 Å |
| $r_{2,3}$ | ([O III]+[O II])/H$\beta$ |

**Table 3.** Key to nomenclature used in the text to identify certain common emission lines produced in the ISM.

To test for parameter sensitivity we varied the relevant parameters one by one, deriving tracks for different ionizing spectra, different internal radii, and different chemical composition, (either the Orion mixture, or a Solar abundance composition). We measured the distances between the tracks in the reference model and the corresponding tracks in the variant models, and also distances between tracks in the different variant models. The results for the FF models and the CLUMPY models are shown in Tables 1 and 2 respectively.

Before commenting on the results of the tests in Tables 1 and 2 we will explain in more detail some of the input data used. To test the effect of a variation in the ionizing spectrum we calculated, in addition to the track obtained with the reference black body at 51230 K, tracks generated with a black body at 41230 K and with O3 stars modelled with the program WM-Basic (Pauldrach et al., 2001). This is a realistic model of the emission from this type of stars, which takes into account line blanketing and the effects of stellar wind production on the final output spectrum. The parameters we need to compute the model output, (stellar temperature, luminosity etc.) were taken from Herrero et al. (2002).

The abundance composition of the ISM is varied by taking two set mixtures of elements: firstly one with solar composition, and secondly one with the composition of the Orion nebula, as given in Ferland et al. (1996). Finally the internal radius of an H II region has to be specified in order not to fall into computational difficulties near the stellar surface. Fortunately the H II regions are obliging in this respect, as the stellar winds from the OB stars do sweep out a low density inner spherical zone. We have specified this in the FF models as between 0.06 pc and 6 pc and we have checked that changing the inner radius over this range does not lead to significant variations in the loci of the tracks. The CLUMPY models are based on a 1 pc characteristic clump size, as observed directly in the dense clumps in the Galaxy ( Cox and Smith, 1974, McKee and Ostriker, 1974, Trapero et al. 1992, 1993, Giammanco et al 2004) and the idea here is that a dense clump of this size will not be too rapidly ablated by the central star, (estimates of the ablation timescales for such clumps can be found in Garcia-Arredondo et al. 2002). The parameters we have selected for our CLUMPY models imply that there will necessarily be at least one clump within 4 pc of the centre of the region, and this is our selected inner radius for these models. We can see in Table 2 that changes absorption factor per spherical shell (which is the equivalent of the number density of clumps) fixed in our CLUMPY models (see Giammanco et al. 2004). Nor have we investigated, in the present section, the effect on a track of changing the input luminosity, since this is one of the free parameters which the grid of models can generate, and which we will be examining explicitly later in the article.

in this value over a small but reasonable range produce very small track shifts. The results of our trials allowed us to concentrate our attention on the most relevant parameters when we came to test the effects of photon escape on the diagnostic diagrams. Firstly we find that in practice a change in the internal radius of the model H II region produces distances between our tracks some two orders of magnitude smaller than those produced by some of the other parameters, as shown in Tables 1 and 2. Thus in practice we are at liberty to select an inner radius from within the ranges tested without significantly affecting any of the diagnostic diagrams. A similar result is found for changes in the global metallicities of the ionizing stars, so that we need not place much emphasis on the precise choice of this parameter. However the input spectral energy distribution and the abundance distribution of the ISM both produce major intertrack distances, which implies that a careful choice of both is very important. Our final choice of spectrum input was made empirically, using spectra from stellar models which reproduce well the "turnover" in the diagnostic diagram for oxygen, as explained in the final paragraph of section 4, below. The same qualitative conclusions are reached from our tests with both FF and CLUMPY models as can be seen from the track separation values in Tables 1 and 2.

As a result of these tests, we opted to use the Orion mixture as the standard, since this is based on observations made in an H II region; we noted that the results obtained using this assumption allowed better general agreement of model predictions and observations than some trials using the solar abundance mixture for the ISM, as we will see in section 4.1 (Fig. 6). As for the input ionizing spectrum our track tests indicated the need for the greatest care in choosing valid input spectra (cf. Fig. 1), a result also reported by Morisset et al. (2004). We have opted for the WM-Basic spectra, as these are based on realistic semi-empirical considerations of the output from hot stars, taking into good account the effects of winds. Within the grid of such spectra we could then use inputs corresponding to the stellar distribution of our choice, as we will see below. For a full discussion of the inclusion of realistic input spectra of ionizing photons in H II regions see Stasińska and Schaerer (1997).

## 3. The effects of a significant escape fraction ionizing photon on estimates the ionization and radiation hardness parameters.

The ionization parameter U is a measure of the ratio of the ionizing photon density to the gas density and, together with the element composition of the gas, determines the emission spectrum of an H II region or diffuse ionized gas region in a galaxy. For this reason it is used as a free parameter in many model grids relating physical parameters to line strengths and ratios. One practical method of determining U was suggested by Díaz et al. (1991): to measure the ratio between the sulphur II doublet [S II] $\lambda\lambda$ 6717,6731 Å and the sulphur III doublet [S III] $\lambda\lambda$ 9069, 9532 Å using the formula:

$$\mathrm{Log}\,(U) = -1.69\,\mathrm{Log}\left(\frac{[\text{S II}]}{[\text{S III}]}\right) - 2.99. \quad (2)$$

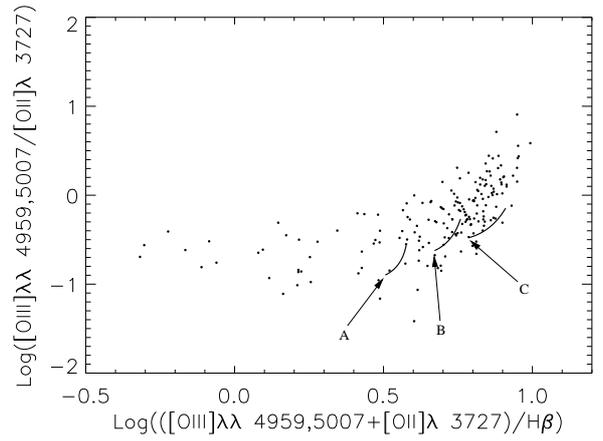

**Fig. 1.** Data points for H II regions from van Zee et al. (1998), plotted in the "oxygen" diagnostic diagram, whose parameters are made up of ratios of line strengths for different ionization states of oxygen (see text for the definition of $r_{2,3}$, which is a ratio including H$\beta$). The three tracks shown are predictions based on CLUMPY models, using different input ionizing spectra: A a black body at 41230K; B a black body at 51230K; C a spectrum defined by the program WM-Basic (for details see the text). We can see that the differences between these tracks are substantial, (see also Table 3). The arrows indicate points of minimum ionizing photon escape from the modelled regions. The range of variation used for $\xi$ in the CLUMPY model is 15%–60%

In this section we will examine the possible effects of photon escape on the determination of U, and for this purpose we have generated three models with different ionizing fluxes for each of our two basic model types: FF and CLUMPY. The ionizing fluxes correspond to 3 O7 stars, to 3 O3 stars and to 30 O3 stars respectively.

In Fig. 2 we show how Log(U), measured using eqn. (2), tends to grow linearly with the escape fraction in conditions of density bounding.
This is an effect that we will observe in all the diagnostic diagrams used in the present paper. It is due to the fact that to allow photons to escape in the models we took gas mass from the HII region by setting a restricted outer radius to the region (we could have used a shell structure with a large inner radius). As the ionizing source remains constant the effect is that the number of ionizing photons divided by the number of H atoms in the region rises as we raise the escape fraction. We note here that the curves for the cases of 3 O7 stars and 3 O3 stars are very closely parallel in both graphs. This implies that any effects of changes in $T_{eff}$ on the resulting value inferred for the escape fraction are of second order in these models.
We can use this result to develop methods of estimating the escape fraction. Using, in the first instance, the assumption of optically thin clumps (the FF assumption) we have an expression for the ionization parameter, $U_\circ$ of an ionization bounded region (for which the escape fraction $\xi=0$) of

$$U_\circ = A\,(Q\,n\,\phi^2)^{1/3}, \quad (3)$$

where A is a known function of temperature (Campbell 1998, Bresolin et al. 1999), and we have $A \sim 2.8\ 10^{-20}\ (10^4/T)^{2/3}$

(Stasińska & Leitherer, 1996). If we use the result implied in Fig. 2 , for a region which has an escape fraction $\xi$ we can write:

$$\mathrm{Log}\,(U) = \mathrm{Log}\,(U_\circ) + m \cdot \xi. \qquad (4)$$

and combining eqns. (4) and (3) we have:

$$\mathrm{Log}\,(U) = \mathrm{Log}\,(A) + \frac{1}{3}\mathrm{Log}\,(Q) + \frac{1}{3}\mathrm{Log}\,(n\,\phi^2) + m \cdot \xi. \qquad (5)$$

In the other hand, the H$\alpha$ luminosity of an H II region (which for convenience here we will call simply H$\alpha$) is given by the product of the number of ionizing photons absorbed and a virtually constant conversion factor, which we term $b$, so that

$$\mathrm{Log}\,(\mathrm{H}\alpha) = \mathrm{Log}\,(\,b\,Q\,(1-\xi)\,). \qquad (6)$$

so that combining equations 5 and 6 gives us:

$$\frac{10^{3m\xi}}{(1-\xi)} = \frac{b}{A^3}\,\frac{U^3}{n\,\phi^2\,\mathrm{H}\alpha}, \qquad (7)$$

On the right hand side of eqn. (7) we find parameters which are either previousl known from physical considerations or measurable from observations, so that if $m$ can be derived we can use 7 to find $\xi$, since the left hand side is a function of only these two variables. We will call this function F($\xi$) and in Fig. 3 we show Log(F($\xi$)) as a function of $\xi$. In order to plot Fig. 3 we first estimated $m$ from Fig. 2, finding a value of 1.5. For a rather wide range of vlaues of $\xi$ we can express the curve in Fig. 3 by Log(F)=5.16$\xi$. So in this range we can write (7) as:

$$5.16\,\xi = 3\,\mathrm{Log}(U) - 3\,\mathrm{Log}(U_\alpha), \qquad (8)$$

were $U_\alpha$ is the ionization parameter estimed using H$\alpha$ luminosity ($U_\alpha = A(n\,\phi^2\,\mathrm{H}\alpha/b)^{1/3}$).

We can now apply the formula in (8) to a set of regions in M 51 studied by Díaz et al. (1991). They derive, for a group of 5 H II regions, the value of $U$ using the formula we give above in (2), the value of $n$ using line ratios, and the number of ionizing photons emitted by their ionizing stars, estimated from the H$\alpha$ luminosity of the regions as given in van der Hulst et al. (1988). For the required steps in the calculation we need the temparature and the filling factor. Díaz et al. (1991) give values for the filling factors, using their derived values for $U$, which would be valid in FF models if the regions are ionization bounded. However here we wish to test these hypotheses. To get a useful starting point, we use a filling factor range $10^{-2}$–$10^{-4}$, which includes the values obtained from direct comparisons of emission measure and local electron density (without the application of $U$) by Rozas et al.(1996) and Relaño et al. (2005), who find values between $10^{-3}$–$10^{-4}$, and for the temperature we use a canonical value of $10^4$ K. Using these values we find that for 4 of the 5 regions tested, the escape fraction of ionizing photons lies between ∼30% and 50%. The fifth region CCM 19 appears to give very low values, and with $\phi = 10^{-2}$ or $10^{-3}$ even a negative value (an impossible result). So most probably it has a rather low value for $\phi$ (we note that Díaz et al., 1991 show this region as having a particularly wide observational error band). In Table 4 we show the results of our "FF" model calculations for the five selected regions, and for completeness we have included three different values of the filling factor, taking as extreme limiting cases $10^{-2}$ and $10^{-4}$.

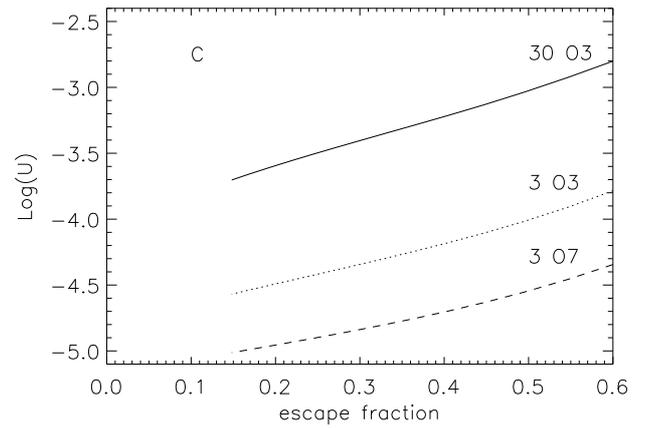

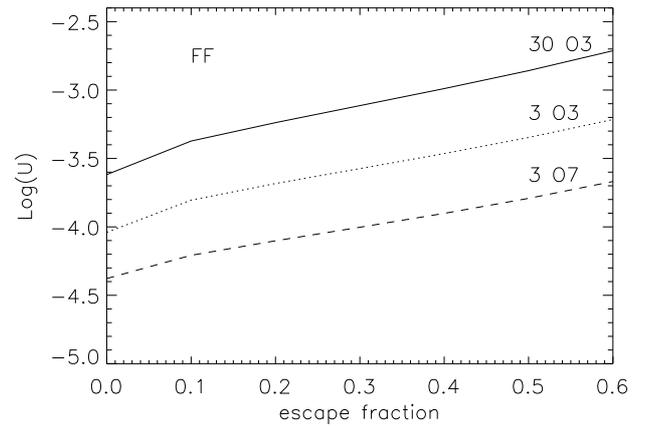

**Fig. 2.** Plots of the variation of the derived apparent value of the ionization parameter $U$ v. the escape fraction $\xi$ of the ionizing photons from an H II region, for CLUMPY models (upper panel) and FF models (lower panel), are for the numbers of equivalent O stars shown. The effect is such that for most diagnostic diagrams a finite value of $\xi$ will be interpreted as an increase in the value of $U$ for the same stellar photon input and gas parameters. The importance of the effect is illustrated by the fact that an H II region with 50% photon escape will show the same value of $U$ as a region with no photon escape and an input ionizing flux almost an order of magnitude greater.

Vílchez & Pagel (1988) introduced the "radiation softness" parameter, $\eta$, which they defined via $\eta=(\mathrm{O}^+/\mathrm{O}^{++})(\mathrm{S}^{++}/\mathrm{S}^+)$, that is an indicator of the effective temperatur of the ionizing source. In order to study it, they define the parameter $\eta\prime$ which is closely related to $\eta$

$$\mathrm{Log}\,(\eta\prime) = \mathrm{Log}\left(\frac{[\mathrm{S\,III}]}{[\mathrm{S\,II}]}\,\frac{[\mathrm{O\,II}]}{[\mathrm{O\,III}]}\right), \qquad (9)$$

and we also have:

$$\mathrm{Log}\,(\eta) = \mathrm{Log}\,(\eta\prime) + \frac{0.14}{t} + 0.16, \qquad (10)$$

were $t$ is the electronic temperaure of the gas normalized for $10^4$.

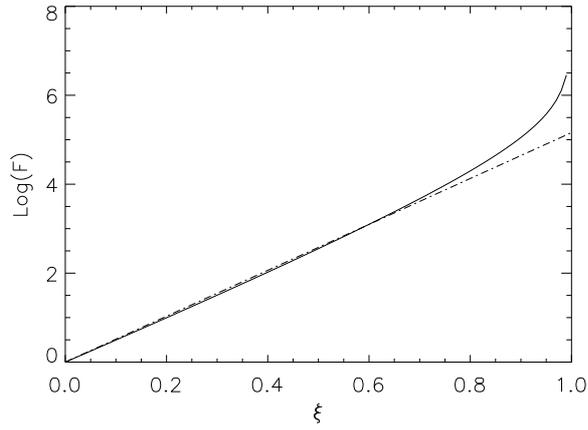

**Fig. 3.** Log of the F($\xi$) as a funciton of $\xi$ (solid line), and the linear relation r=5.16·$\xi$ (dotted-dashed line), showing that F($\xi$) behaves quasi-linearly for a wide range of $\xi$ values. The value of F can be derived from a weighted ratio of the ionization parameter, $U$, and the absolute H$\alpha$ luminosity of a given H II region.

| Region | CCM 72 | CCM 24 | CCM 71 | CCM 10 | CCM 19 |
|---|---|---|---|---|---|
| Log(Q) | 51.70 | 51.14 | 51.27 | 51.51 | 51.00 |
| Log(U) | -3.10 | -3.00 | -3.00 | -3.05 | -4.00 |
| $\xi_{\phi=10^{-2}}$ | -6% | 10% | 8% | -0.2% | -45% |
| $\xi_{\phi=10^{-3}}$ | **32%** | **48%** | **46%** | **38%** | **-6%** |
| $\xi_{\phi=10^{-4}}$ | 70% | 87% | 85% | 77% | 32% |

**Table 4.** Estimated values of escape fraction for the H II regions presented in Díaz et al. (1991), from where we have taken the nomenclature for the regions. We take three values for $\phi$ to bracket the range $\sim 10^{-3}$–$10^{-4}$ infered observationally by Rozas et al. (1996) and Relaño et al. (2005) the results for this value are in bold face. Where the sign is negative the assumed value of $\phi$ or the infered value of $U$ cannot be correct.

There are a numbre of studies which give as a result a subsidiary dependence of $\eta$ on the metallicity of an H II region (Stasińska 1980, Campbell et al. 1986, Melnick, 1992, Cerviño & Mas-Hesse 1992, Bresolin et al. 1999, Cedrés & Cepa 2002). In Fig. 4 we show that $\eta\prime$ varies not only with the metallicity but also with the escape fraction $\xi$, in a rather simple way for CLUMPY models but in a rather more complex way for FF models. We need not describe the complexities in more detail, but it is reasonable that the behaviour of $\eta$ is different, for the two types of models, because $\eta$ is a measure of the ratio between the flux of high energy photons (capable of ionizing He), and the remainder of the photon spectrum. The way that a model of the ISM absorbs differentially these two parts of the spectrum will clearly be different for models with optically thick an optically thin clumps. Essentially in a CLUMPY model a clump either absorbs the full spectral range or transmits it, whereas in an FF model there is a differential effect favouring the transmission of the harder photons through the optically thin clumps, so the behaviour of $\eta\prime$ with $\xi$ will not be the same. In spite of this, however, in all our models the variations introduced by allowing a fraction of the photons to escape are of the same order as those produced by reasonable variations in the metallicity. This means that in order to obtain a reliable value for $\eta\prime$ it is important to make the best possible estimate of $\xi$, the photon escape fraction.

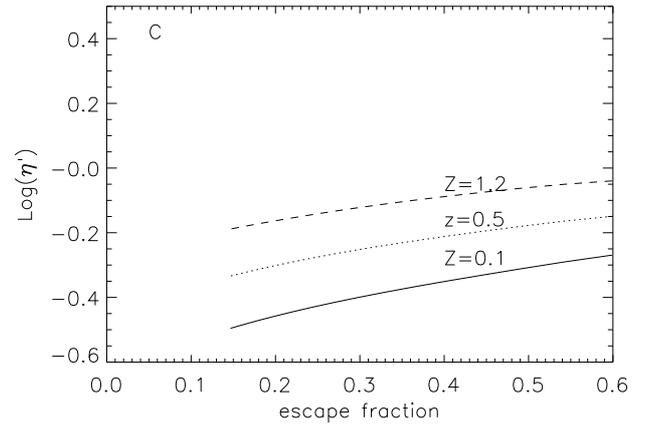

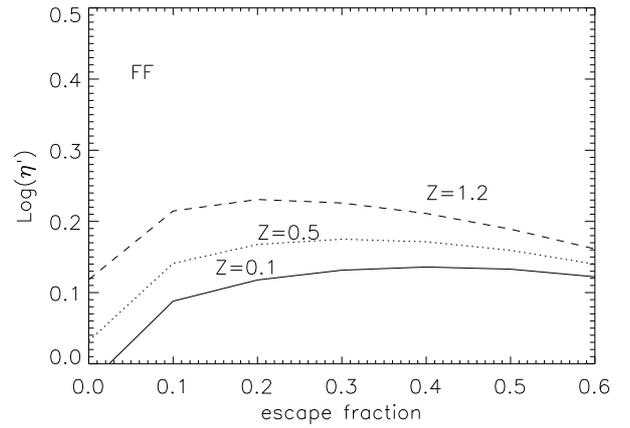

**Fig. 4.** Log of the derived spectral hardness parameter $\eta\prime$ v. the escape fraction of ionizing photons $\xi$. The models are for a specific photon input, that due to 3 O3 stars, and with three different metallicities for the gas. This figure shows the considerable differences in $\eta\prime$ which occur if the assumptions of CLUMPY (upper panel) or FF (lower panel) models are taken.

## 4. Effects in oxygen line diagram

In section 3 we saw how density bounding in an H II region can mimic a variation in the ionization parameter. This variable, the ionization parameter is currently used as an independent variable when generating predictive model grids in diagnostic diagrams, so that the overall morpholgical features of these grids should not vary greatly if the hypothesis of density bounding is adopted, in fact substantive changes do originate only if we assume that the medium has optically thick clumps (i.e. in our CLUMPY models). It is these changes of form and their implications which we will describe in detail in the present section.

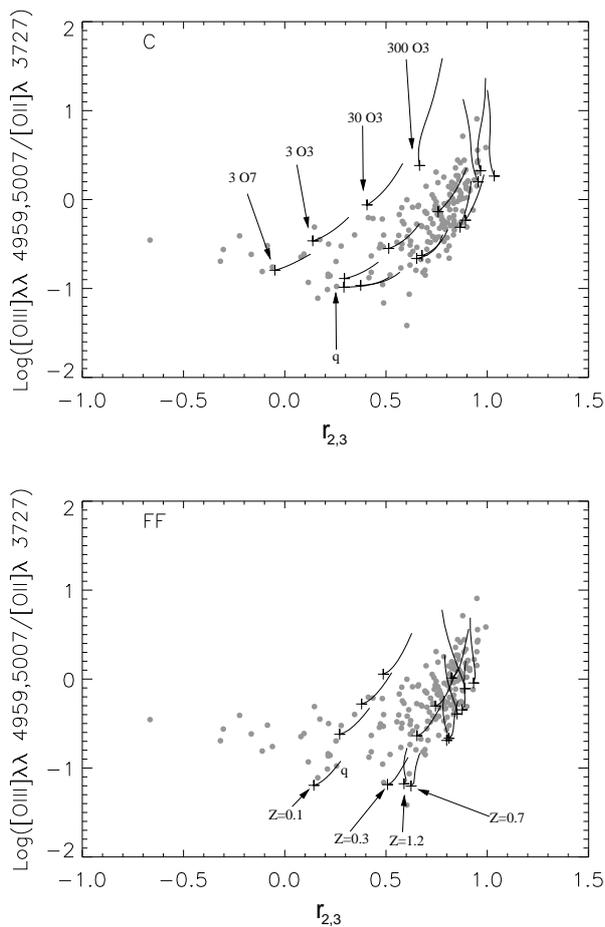

**Fig. 5.** Observational results v. model predictions for the "oxygen" diagnostic diagram. The points are derived directly from observations (references in fig. 1). In the upper panel we show theoretical tracks produced by CLUMPY models and in the lower panel traces from FF models. Each track in both diagrams corresponds to a set of model predictions using photon inputs from the equivalent number of O stars labelled in the upper diagram, and of gas metallicities labelled in the lower diagram. Each track has minimum photon escape at the bottom (0% for FF models, 15% for CLUMPY models) and maximum photon escape (60% for both). at the top. For the FF models joining the crosses for zero photon escape will reproduce the predictions of standard ionization bounded models. The diagrams are intended to show the general trends in the model predictions rather than to pick out results for a given metallicity or ionizing luminosity. The test point q, as explained in the text, illustrates the model dependence of the inference of abundances from emission line data.

In Fig. 5 we show a diagram which has been frequently used (see e.g. McCall et al., 1985), it is the plot of Log([O III]/[O II]) v. $r_{2,3}$ for both FF and CLUMPY models where a set of observational points (van Zee et al., 1998) is compared with prediction for both FF and CLUMPY models. We can see at once that the overall morphology of the model tracks depends quite strongly on the type of models considered. This is illustrated by the point 'q' in Fig. 5 which for the CLUMPY tracks is found between metallicities 0.3 and 0.5 while for the FF tracks it is situated at metallicity 0.1 (note that we have defined 'Z', the metallicity, as a factor wich multiplies the mean abundance of metals in the Orion mixture). This shows that the use of this diagram to derive metallicities can be seriously model dependent, and it is necessary to find methods which enable us to decide between model types.

In order to attempt to quantify the difference between a data set and a global set of model predictions we have defined the quadratic distance $\Delta^2$ between a data set and a model grid as the sum of the least squared distance values between data points $i$ and models $j$. Written formally this is:

$$\Delta^2 = \Sigma_i \ min \ (d^2_{i, \ j=1,2...j\neq i}) \qquad (11)$$

If $\Delta^2$ were to take the value zero this would mean that the grid coincides exactly with the data points, and the larger the value of $\Delta^2$ the further are the points from the grid.

In the diagnostic diagram in Fig. 5 we find values for $\Delta^2$ of 3.2 using the grid of FF models and 1.9 using the CLUMPY model grid, which indicate that the CLUMPY models are giving a better account of the data, as the steps of the grids are the same in both cases. This result taken alone is not more than a pointer. Indeed if the data were uniformly distributed in parameter space, and the set observed were representative, $\Delta^2$ should have a precise statistical meaning. However neither of these conditions is realized, so we will not deepen our discussion of $\Delta^2$. Even so it is of interest that the conclusions reached using $\Delta^2$ agree with a visual appraisal. We observe that the models predict a "forbiden zone", and therefore a "permitted" zone: in the CLUMPY case the permitted zone shows a morphology closer to that of the data, than in the FF case.

The existence of a forbidden zone in the oxygen line diagram is a consequence of the 'turnover' in metallicity of the H II regions, such that as the metallicity Z rises the parameter $r_{23}$ also rises but reaches a maximum value and then declines (see e.g. Pilyugin, 2000 for illustrations of this). In our model (FF and CLUMPY) the turnover is at Z = 0.7 which is equivalent to Log(O/H) = -3.6. This value is in agreement with that obtained by McGaugh(1991). In Fig. 5 we can see that the edge of the forbidden zone predicted by the models coincides with the observational edge. If we alter the ionizing spectrum the boundary of the permitted zone is shifted as can be seen in Fig. 6 where we have produced theoretical tracks using as input a black-body at 51230 K. We can see that the theoretical turnover, the locus of points where the tracks stop moving to the right in the diagram and begin to move back to the left, is now well to the left of the boundary of the observations, so that this input spectrum cannot yield the observational points for any metallicity. We do note, however, that the turnover occurs at the same metallicity as before, Log([O/H]) = -3.6. Finally, as the WM-Basic stellar models do produce a turnover at the boundary of the observations, this gives us a valid reason to use these stellar atmospheres in the rest of our study.

## 5. Pilyugin's data compared with model prediction

One of the principal ways of determining element abundances in extragalactic H II regions is empirically via the parameter

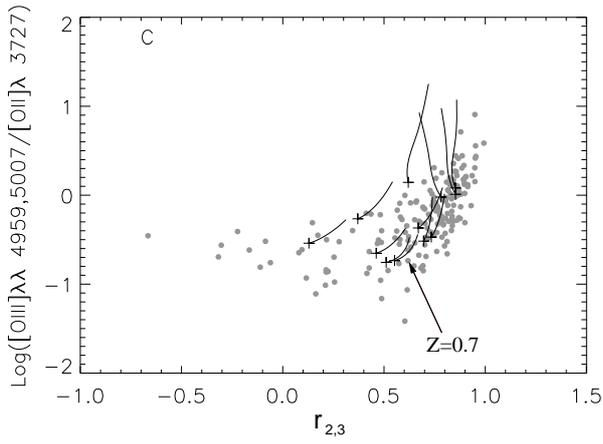

**Fig. 6.** "Oxygen" diagnostic diagram using the same parameters as in Fig. 5 for CLUMPY models, but with a black-body input spectrum instead of the more realistic stellar spectral inputs given by CLOUDY. The zone of the turnover, where the predicted values of $r_{2,3}$ stop rising and move downwards with increasing metallicity, is clearly displaced from the data in these models, though the value of [O/H] at the turnover, -3.6, is the same as in the better fitting models shown in Fig. 5. Using the Solar and Orion abundances set we finde a simila beaviour. The Orion set give a better fit of the data. We remember in any case that the term Solar indicate the Ferland (1996) selection the now is quite old. The new estimation of oxygen abundances is more similar to the Orion mixture (Allende Prieto et al. 2001)

$r_{2,3}$, using a calibration relation between the oxygen abundance and $r_{2,3}$, whose data points come from a set of Galactic H II regions where it is possible to measure the abundance more directly by althernative methods (see Pilyugin, 2000). With the aim of comparing our models with a more ample set of data points, we present our own plots of the oxygen abundance v. $r_{2,3}$ where the tracks are computed using either CLUMPY models (Fig. 7, upper panel) or FF models (Fig. 7 lower panel). We can see that for both types of models the distribution in the parameter plane is reasonably well reproduced by the model tracks (compare these figures with Pilyugin, 2000, Fig. 1). Specifically the scatter in the $r_{2,3}$ coordinate is consistent with variations in U as the cause.

We have seen previously in section 3 that a rise in the escape fraction of ionizing photons from an H II region causes a rise in the estimated value of U. In fact we saw that U depends on Q, $n$, $\phi$ and (at constant gas temperature) is also a rising function of $\xi$. So in any diagnostic diagram an increase in escape fraction could (and typically would) have been interpreted as a rise in the ionizing flux per unit gas mass, always assuming that $n\phi^2$ is invariant. In the next section we will discuss this specific degeneracy, and attempt to remove it via a further diagnostic.

## 6. Diagnostics using absolute fluxes

In the previous sections we have seen how photon flux escape can be intepreted in the majority of the diagnostic diagrams, normally employed, as a rise in the ionizing parameter $U$, and hence as a rise in the ionizing flux, if $n\phi^2$ remains essentially constant.

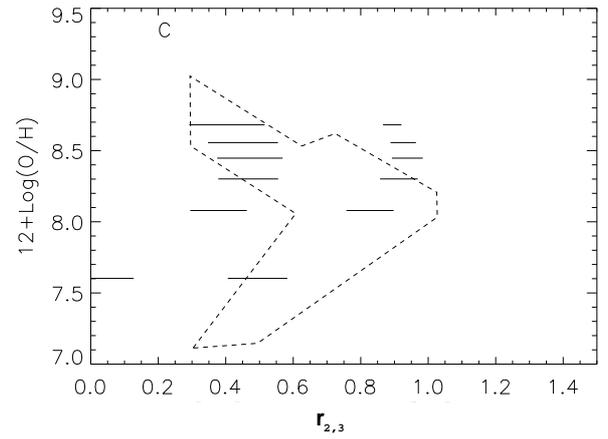

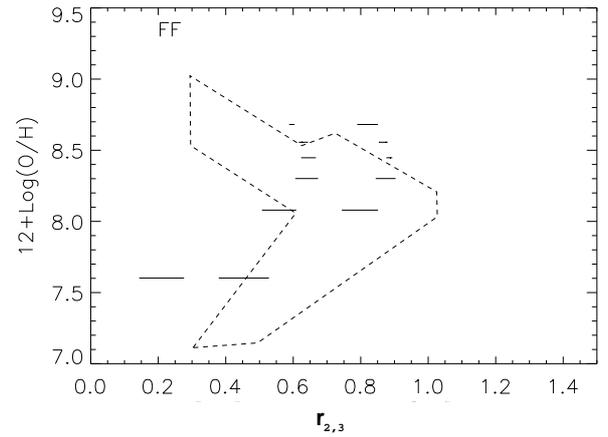

**Fig. 7.** Oxygen abundance v. the $r_{2,3}$ parameter for H II regions traks (the orizontal bars). The upper panel is for CLUMPY models, and the lower panel for FF models. In both diagrams, the tracks at the left correspond to a set of model predictions using photon inputs from the equivalent number of 3 O7 stars, the traks at the right have been obtained using 30 O3 equivalent stars. The region marked from a discontinuus line represente de loci of Pilyugin data (Pyliugin, 2000 fig. 1). In the traks the photon escape comes from a minimum in the left to a maximum in the right. The orizontal spread in the data should be interpreted as a variation of ionization parameter $U$, or as a variation of photon escape fraction.

We could in principle eliminate this degeneracy by studying the distribution of H II regions in the Log ($U$)-Log (H$\alpha$) plane. For the moment we do not have satisfactory sets of relevant data for H II regions, in this plane because it is difficult to find measures of [S II] and [S III] from which to estimate the ionizing parameter[1].

In the absence of such measures of the ioning parameter, we can use the [O III]/[O II] to obtain a first approximate estimate of U, and in this case we do have a limited data set where this ratio is obtained in H II regions whose H$\alpha$ luminosity has been measured.

---

[1] We should note that although eqn.7 holds only for FF models, in the CLUMPY models there is always significant photon escape, so it is not necessary to devise a specific test in this case.

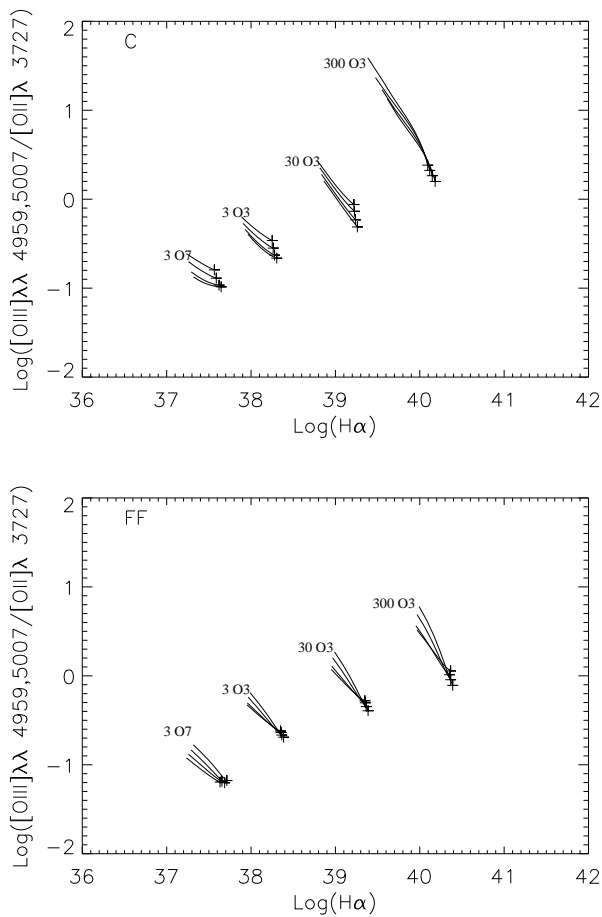

**Fig. 8.** Diagnostic diagram for photon escape obtained using CLUMPY models, (upper panel) and FF models (lower panel). Each set of tracks is a prediction of the [O III], [O II] line ratio for the intrinsic ionizing input flux implied by the number of equivalent O stars indicated. Each cluster of tracks is the result of varying the metallicity of the gas, and we can see that this has a relatively small effect on the ratio. The ionization parameter $U$ is classically estimated via this ratio, assuming no photon escape. However an artificially high value of $U$ due to photon escape will be indicated by a low H$\alpha$ luminosity, and this can be quantified (see text).

In Fig. 8 we show how the tracks in this parameter plane are predicted with both FF and CLUMPY models. The degeneracy between $\xi$ and Q is in fact resolved, as the tracks corresponding to different input ionizing luminosities are now separated in the parameter plane. In practice an increase in the escape factor does bring about an increase in U, but reduces the H$\alpha$ luminosity so that the tracks due to different ionizing luminosities are not superposed.

For the FF models the ionization bounded points (i.e. those with zero photon escape) lie on virtually a straight line, below which the tracks run out, and above which the regions are density bounded. For the CLUMPY models the situation is rather similar, although here there is a lower limiting escape fraction of 15%. The models with lowish escape fraction lie in a band rather than along a line, since models with different metallicities but identical input ionizing fluxes yield slightly different line ratios. However the scatter induced by variations in the photon escape fraction is much greater than the width of this band.

We will now take a brief analytical look at a data set of line strengths obtained for H II regions in M 101 using the oxygen diagnostic diagram (cf. Fig. 5) and also the photon escape diagram (cf. Fig. 8). The data for M 101 was extracted from Cedrés & Cepa (2002). The acquisition and reduction of the data are explained there. The data includes extinction corrected fluxes for 338 regions in the emission lines: H$\alpha$, H$\beta$, [O II], [O III] and their respective continua. The accuracy and reliability of this data is shown in Cedrés et al. (2004), where the oxygen abundance for the inner parts of M101 is determined employing Pilyugin (2001) calibration, and compared with the abundance obtained by Kennicutt et al. (2003).

In Fig. 9 (upper panel) we have marked a set of data points with arrows. These are regions which in the oxygen line diagram occupy the zone corresponding to models generated by stars with 30 or fewer O3 stars, but with a high escape fraction, or alternatively by regions which are ionization bounded (zero escape) but which contain between 100 and 300 O3 stars. These are clearly regions with extreme parameters, for which it is relatively easy to discriminate valid from invalid models, as it is improbable that the ionizing star clusters are so hugely luminous, so it is much more likely that they show considerable photon escape. For comparison we show the same points in the "escape" diagram Fig. 9 (lower panel) and as we would have predicted they lie in a zone occupied by regions with high values of $\xi$. We must also note that the scatter of the points is greater than that predicted by our models in general, which could be explained by a combination of the usual observational errors in deriving line ratios and fluxes, plus the effects of varying H II region morphology (our models here assume spherical symmetry, and any departure from this towards more realistic structure would lead to increased photon losses for real regions). In a more complete treatment we should take into account variations in $n\phi^2$, but for the time being the required complete data sets of line diagnostics to determine this in the presence of photon escape are not available.

## 7. Conclusions

In this paper we have considered how model H II regions with inhomogeneities can give a particularly good account of the emission line intensities and ratios observed in real regions. We have considered two different types of models: those with optically thin density clumps, termed 'FF' models because they are the equivalent of the classical H II region models with dense material present in a fraction of the total volume, the "filling factor", but with clumps small enough to be completely ionized, and those with optically thick density clumps, which we have termed 'CLUMPY'.

In order to optimize the comparison of models with observa-tions we have defined the concept of a track in a diagnostic diagram as the ensemble of models where only one parameter is varied, and we have used the escape fraction of ionizing photons from the regions as the parameter to vary. We have used diagnostic diagrams with a consistent record of use in the literature (oxygen line diagram). We find, quite generally, that the effect of an increasing ionizing photon escape fraction is equivalent to that of an increasing ionization parameter so that the hypothesis that H II regions are all, to a lesser or greater degree, leaky is compatible with the observations.

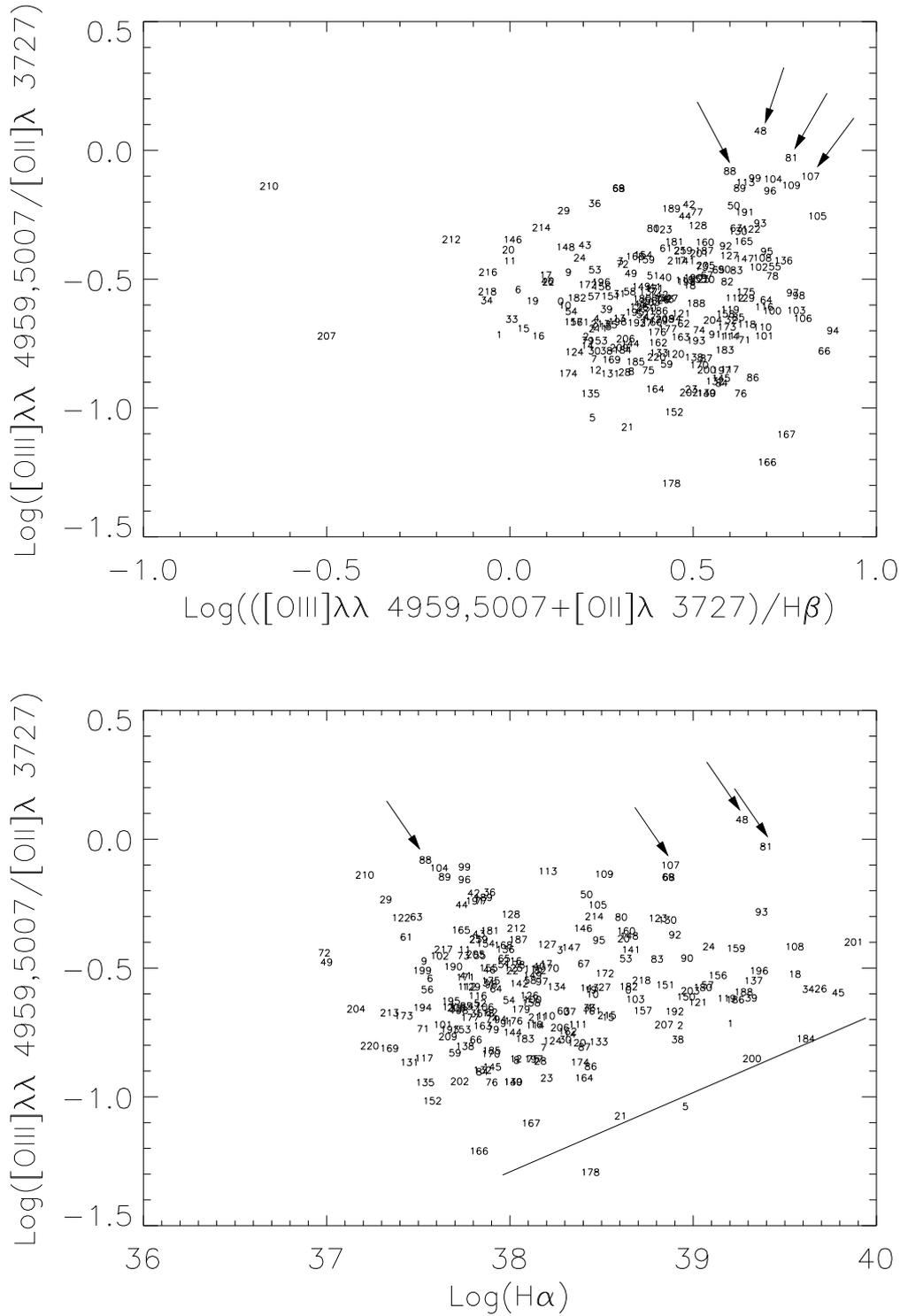

**Fig. 9.** Upper panel. Data points for H II regions in the galaxy M 101, from Cedres & Cepa (2002) in the oxygen diagnostic diagram. Using the models illustrated in Fig.5 we can show that this zone of the diagram is populated by H II regions which, at one extreme, contain more than 100 O3 stars with no photon escape, and at the other extreme contain some 30 O3 stars but with ∼ 60% photon escape. Lower panel. Here we have plotted [O III]/[O II] line intensities v. the absolute H$\alpha$ luminosities for the same set of regions. Comparing this plot with Fig.5 we can see that the arrows in fact correspond to regions with major photon escape fractions. It is also clear that the majority of the H II regions observed appear to have lower photon escape fractions. The solid line is the locus of zero predicted photon escape; the locations of the data points with respect to this locus implies that virtually all the H II regions observed do show significant photon escape.

However the object of our exercise is, or should be, to estimate the escape fraction of the ionizing photons. We have resolved this using a semi-analytic method, and produce an expression which allows us to estimate escape fractions in real cases using measurable or known quantities. We have used this to make estimates for selected observed H II regions in M 51, finding characteristic values of between a third and a half for the escape fraction.

Our diagnostic scheme contains an intrinsic degeneracy: that for increasing fractional photon escape the ionization parameter derived using the usual line ratio methods will be larger, and we may have no a priori knowledge of which of the two effects we are measuring. However we can allow for this, because for a higher escape fraction the resulting H$\alpha$ luminosity of the H II region will be reduced, so that if an estimator of the ionization parameter is plotted against H$\alpha$ luminosity the degeneracy can be resolved. Using this technique we show, for example, that there are extremely clear cases of ionizing photon escape for an identifiable subset of H II regions in M 101. In principle we can obtain an absolute calibration for this graph in terms of the ionization parameter. To do this we need to find one or more regions for which we can derive the absolute escape fraction, e.g. by using equation (7), or more reliably using an H II region sufficiently local that a Lyman continuum budget can be produced on the basis of a comparison of emission lines with the known outputs of the ionizing stars, as in Relaño et al. (2002).

In this paper we have verified that the emission line ratio data in the literature are compatible with the hypothesis that H II regions are leaky, and have gone a step further, showing that for many specific cases the most satisfactory interpretation of line ratios and intensities is obtained by assuming photon escape. This general inference is valid for both types of models (both 'FF' and 'CLUMPY') although the detailed derivation of parameters will, in general, differ from one type of models to the other. We have proposed a way to quantify the escape fraction using only observed line ratios and absolute fluxes from the H II regions themselves, without necessarily refering to measurements of these parameters in the surrounding diffuse medium, though this does offer a valuable check. We conclude by admitting that it is not generally easy in practice to tell observationally whether or not the CLUMPY assumption gives a better account of the measurements than the FF assumption, but that this degree of uncertainty depends largely on lack of the relevant complete data sets of line ratios and also absolute luminosities, which is a deficiency which can in principle be remedied in a straightforward way.

*Acknowledgements.* We are happy to thank Almudena Zurita for valuable discussions and Jordi Cepa for providing observational data. This work was supported by the Spanish Ministry of Science and Technolgy under grant AYA2002-01379 and by the Ministry of Education and Science via grant AYA 2004–08251–C02–01 The CLOUDY suite is available for use, thanks to its authors, Dr. G. Ferland and collaborators. We are happy to thank the anonymous referee for useful comments which have helped us to improve the paper and its presentation.